\documentstyle[prb,aps,12pt]{revtex}
\begin{document}
\draft
\title{Solitons and (spin-)Peierls transition in disordered
quasi-one-dimensional systems}
\author{M. Mostovoy\footnote{also at G. I. Budker Institute of
Nuclear Physics, 630090 Novosibirsk, Russia.} and D.
Khomskii\footnote{also at P. N. Lebedev Physical Institute, Leninski
prosp. 53, Moscow, Russia.}}
\address{University of Groningen, Nijenborgh 4, 9747 AG
Groningen, The Netherlands}
\date{\today}
\maketitle
\begin{abstract}
\baselineskip=18pt
\widetext
\leftskip 54.8pt
\rightskip 54.8pt
We study the (spin-)Peierls transition in quasi-one-dimensional
disordered systems, treating the lattice classically.  The role
of kinks, induced thermally and by disorder, is emphasized.
For weak interchain interaction the kinks destroy the coherence
between different chains at a temperature significantly lower
than the mean-field Peierls transition temperature.  We formulate
the effective Ising model, which describes such a transition,
investigate the doping dependence of the (spin-)Peierls
transition temperature and discuss several implications of the
picture developed.  The results are compared with the properties
of the spin-Peierls system CuGeO$_3$.

\end{abstract}


\baselineskip=18pt

\section{Introduction}

The tendency towards the lattice instability is strongly enhanced
in quasi-one-dimensional systems.  Conducting chains are
inherently unstable against the periodic lattice distortion
which opens a gap in the spectrum of electron excitations
(Peierls instability).\cite{Peierls} TaS$_3$, NbSe$_3$,
K$_{0.3}$MoO$_3$ are the examples of the quasi-one-dimensional
conductors, which at a sufficiently low temperature become
Peierls insulators.\cite{1d} The periodic distortion was also
observed in organic (TTF-CuBDT \cite{Bray}, TTF-AuBDT
\cite{Jacobs}, (MEM)-(TCNQ)$_2$ \cite{Huizinga}, {\em etc.}) and
inorganic (CuGeO$_3$ \cite{Hase}, NaV$_2$O$_5$
\cite{Isobe,Fujii}) compounds, consisting of weakly interacting
spin chains.

The excitation spectrum in the quasi-one-dimensional electronic
and spin Peierls materials is usually interpreted in terms of the
excitations in isolated chains.  Nonetheless, the (spin-)Peierls
transition is a three-dimensional phenomenon, because
electronic or spin chains are immersed in a crystalline lattice
of a bulk material.  In the standard description of the
(spin-)Peierls transition the lattice is treated in the
mean-field approximation.\cite{RS,Pytte,BBK,CF} In this approach,
the excitations, which eventually destroy the ordered Peierls
phase at a critical temperature, are the electron-hole (spin)
excitations in isolated chains with a fixed lattice distortion.

It is well-known, however, that in these systems there exist also
excitations of a different nature: solitons, or kinks, which are
the mobile domain walls.  In this contribution we consider the
Peierls systems, in which the lattice in the ordered phase is
dimerized, {\em i.e.} long and short bonds alternate.  In this
case the distortion wave is commensurate with lattice and cannot
slide as a whole.  On the other hand, kinks, separating
domains with the opposite sign of the dimerization, can propagate
along chains.  Solitons are mixed electron-lattice (or
spin-lattice) excitations.  Their topological nature is
responsible for the anomalous charge-to-spin ratio, which was
used to explain the unusual bevavior upon doping of the
electronic Peierls material trans-polyacetylene.\cite{SSH} For
spin-Peierls systems, the soliton picture is much less
popular, although it was discussed also in this context in
Ref.\CITE{NF}.

In addition to kinks, excited thermally or optically, there can
be also kinks induced by disorder.  Here we consider only one
kind of disorder, which effectively cuts chains into finite
segments.  The interruption of chains may be caused by strong
conformational disorder ({\em e.g.} bond bending in
trans-polyacetylene), or by substitution of atoms in chains (as
in CuGeO$_3$).  The minimal energy lattice configuration strongly
depends on whether the number of atoms in a chain is even or odd:
an odd chain contains one kink,\cite{Su,Cam} while an even chain
does not.  The disorder-induced kinks are neutral and have spin
$\frac{1}{2}$.  In fact, any kind of the off-diagonal disorder
results in the appearance of kinks,\cite{Iwan,MFK} so the picture
we consider here is rather general.

Although kinks provide a coherent explanation of a vast variety
of experimental data on optical and magnetic properties of
trans-polyacetylene,\cite{SSH,Kies} their role in the Peierls
phase transition was not properly studied.  That may be related
to the fact, that trans-polyacetylene has a large gap $E_g
\approx 1.8$eV, so that the transition into the undimerized state
cannot be observed.  On the other hand, the critical temperature
in spin-Peierls materials is rather low (2--35\,K), and there
exist a very detailed information on these transitions.  An
important property of the recently discovered inorganic
spin-Peierls compound CuGeO$_3$ is its ability to accept dopands,
so that the disorder can be introduced into this system in a
controlled way.  This made possible the systematic study of the
effect of disorder on the spin-Peierls
transition.\cite{Zn1,Zn2,Zn3,Si1,Si2} Finite chains in the
spin-Peierls compound CuGeO$_3$ are formed upon the substitution
of Cu atoms in the CuO$_2$ spin chains by nonmagnetic Zn, which
interrupts the spin-exchange.  Similar effect is reached by the
substitution of Ge by Si.\cite{KGM} Doping by Zn or Si strongly
decreases the Peierls transition temperature and gives rise to
the antiferromagnetic phase,\cite{Zn2,Si1} which coexists with
the dimerization.\cite{Zn2,Si2} We will show, that these effects
can be naturally explained in terms of the disorder-induced
kinks.

In this contribution we consider the (spin-)Peierls transition
into a dimerized state in disordered quasi-one-dimensional
systems, taking into account the soliton excitations.  We believe
that, due to their topological nature, kinks play an important
role in the description of this transition, even though the
interaction between the chains can considerably modify their
properties.  The approach we propose is, in a sense, opposite to
the mean-field description of the (spin-)Peierls transition: we
neglect the electron(spin) excitations, as well as the small
fluctuations of the order parameter, and leave only kinks.  In
other words, we assume that locally the dimerization persists
well above the critical temperature, and that the transition
occurs when the coherence between the dimerization phases on
different chains is established.  The effective model, which
describes such transition in clean systems, is the strongly
anisotropic Ising model.  We propose a simple modification of
this model, which allows us to consider both thermally and
disorder-induced kinks.

Our approach is valid if the coherence of the (spin-)Peierls
state is destroyed by the kinks at a temperature significantly
lower than the mean field transition temperature, so that the
gap in the spectrum of the (spin) electron excitations is still
large at the phase transition point and all nontopological
excitations can be neglected.  As we shall see, this requires a
rather weak interchain interaction.  Another important
approximation we make, is the classical treatment of the soliton
motion.  The classical treatment of lattice is justified, if
the value of the relevant phonon frequency is much smaller than
the value of the Peierls gap.  This is usually the case for the
electronic Peierls materials ({\em e.g.} trans-polyacetylene).
On the other hand, in spin-Peierls materials with a
relatively small spin gap this would require a
particularly soft lattice, which, in principle, can happen due to
a proximity to some structural transition.\cite{BBK} If, on the
contrary, the Peierls gap is comparable to the optical phonon
frequency, or is much smaller, then the consideration of the
three-dimensional lattice has to be quantum.

\section{Weakly coupled finite chains}

In this section we introduce a simple model of disordered
quasi-one-dimensional (spin-)Peierls material.  There are
three essential assumptions on which our model is based: (i)
interchain interaction is assumed to be sufficiently weak; (ii)
off-diagonal disorder is assumed to break effectively
chains into segments; (iii) lattice can be treated
classically.

Because of the weakness of the interchain interaction, in the
first approximation chains can be considered independently of
each other.  The typical length of a segment in a disordered system
is inversely proportional to the concentration of impurities.
The classical lattice configuration of a finite chain has to be
determined by minimizing the total chain energy, which includes
the electron energy (the energy of the spin system for
spin-Peierls materials), as well as the lattice energy.  Below we
discuss the form of the minimal energy lattice configuration of
finite chains.

We begin by considering a conducting chain described by
the Peierls-Hubbard (PH) Hamiltonian,
\begin{eqnarray}
\label{PH}
H_{PH} = &-& \sum_{n,\sigma}\left(t_0+\alpha(u_n-u_{n+1})\right)
(c^{\dagger}_{n\sigma}c_{n+1\,\sigma}
+c^{\dagger}_{n+1\,\sigma}c_{n\sigma}) +
U \sum_n c^{\dagger}_{n\uparrow}c_{n\uparrow}
c^{\dagger}_{n\downarrow}c_{n\downarrow}\nonumber \\
&+& \frac{K}{2}\sum_n(u_n-u_{n+1})^2\;,
\end{eqnarray}
where the operator $c_{n\sigma}$ annihilates an electron with
spin projection $\sigma$ at a site $n$, and $u_n$ is the shift of
the $n$-th atom in the chain direction.  The energy of a
half-filled chain with even number of atoms and periodic boundary
conditions is minimal when the chain is uniformly dimerized,
\begin{equation}
u_n = (-)^n u_0\;,
\end{equation}
which corresponds to alternation of long and short bonds.  The
dimerization takes place at all values of $U$.  For
non-interacting electrons ($U = 0$),\cite{SSH,TLM}
\begin{equation}
\frac{\alpha u_0}{t_0}
\propto e^{- \frac{1}{\lambda_{ep}}}\;,
\end{equation}
where $\lambda_{ep}$ is
the dimensionless electron-phonon coupling constant,
\begin{equation}
\lambda_{ep} = \frac{4 \alpha^2}{\pi t_0 K}\;.
\end{equation}
In
the opposite limit of strong electron correlations, $U \gg 4
t_0$, the value of the dimerization is determined by the
interaction of the lattice with the low energy spin degrees of
freedom, described by the spin-Peierls (SP) Hamiltonian,
\begin{equation}
\label{SP} H_{sp}=\frac{1}{2}\sum_n
\left(J_0+\alpha^{\prime}(u_n-u_{n+1})\right)
(\vec S_n\cdot\vec S_{n+1})
+\frac{K}{2}\sum_n(u_n-u_{n+1})^2\;,
\end{equation} where
\begin{equation}
J_0 = \frac{4 t_0^2}{U}\;,\;\mbox{and}\;
\alpha^{\prime} = \frac{8 t_0}{U} \alpha \;.
\end{equation}
In this case,\cite{CF}
\begin{equation}
\frac{\alpha u_0}{t_0} \propto
\left( \frac{t_0 \lambda_{ep}}{U} \right)^{\frac{3}{2}}\;.
\end{equation}

The number of bonds in even chain with periodic boundary
conditions (a ring) is also even, so that there are two
degenerate minimal energy lattice configurations, shifted by one
lattice constant with respect to each other.  The amplitude of
the distortion wave, $u_0$, for these two lattice configurations
has opposite signs.  The boundary conditions for finite chains
produced by substitution of atoms are, however, free rather than
periodic.  For an open chain with even number of atoms there is
only one minimal energy configuration, since the number of bonds
in such chain is odd, so that the configuration, which has more
short bonds, has lower energy (see Fig.1a).  The value of the
dimerization changes somewhat from the middle to the ends of the
chain, but the sign of the dimerization is everywhere the same.

On the other hand, if the number of atoms in a chain is odd, the
minimal energy lattice configuration has a form of a kink (see
Fig.1b).  The chain energy is practically independent of the
position of the kink, provided that the distance from the kink to
the nearest chain end is larger than the kink size.  One unpaired
spin is localized near the kink.  Therefore, a spin $\frac12$
object in an isolated odd segment, contrary to a common belief,
is localized not close to an impurity (chain end), but rather
away from it.

Next we consider the effects of interchain interaction.  This
interaction may originate, for instance, from a relatively small
hopping of electrons between chains (which corresponds to the
exchange between spins sitting on neighbouring chains in
spin-Peierls materials), long range Coulomb interaction between
the electrons and deformations of the three-dimensional lattice.
Independently of the origin, the interchain interaction tends to
create a coherence between the phases of the distortion waves on
different chains.  In a clean material it determines a type of
crystalline structure.  For simplicity we will assume, that the
lattice configuration has the lowest energy when the sign of the
dimerization is the same on all chains.  (Actually, one can show
that in CuGeO$_3$ the distortions in neighbouring chains along
$b$-axis have opposite phases, but it does not modify our
conclusions.)

When a finite chain, created by disorder, is surrounded by other
chains, its lattice configuration is different from that of an
isolated chain.  For instance, there appear two kinds of even
chains: those in which the sign of the dimerization, favoured by
the boundary conditions, is the same as the sign of the average
dimerization in the bulk material, and those in which it is the
opposite (see Fig.2a).  In the latter case the energy loss due to
interchain interaction is proportional to the chain length.
Therefore, despite the weakness of interchain interaction it
will be energetically more favourable in a sufficiently long
chain to create a kink near one end of the chain and antikink
near the opposite end, so that the sign of the dimerization
between the domain walls is the same as the sign of the
averaged dimerization (see Fig.2b).  Similarly, the interchain
interaction binds kink in an odd segment to one of the chain ends
(see Fig.3).  This, in a sense, restores the conventional picture
of a spin $\frac12$ sitting relatively close to impurity at an
average distance of the order of a kink size $\xi$ (see
also Ref.\CITE{KGM}).

\section{Qualitative discussion of the phase transition in clean
and disordered systems}

In this section we discuss qualitatively the (spin-) Peierls
phase transition in quasi-one-dimensional systems with very weak
interchain interaction.  First we consider the clean systems.
The long range order in an infinitely long isolated chain is
destroyed by thermally excited kinks at any non-zero temperature.
This happens because the energy of kink does not depend on its
position in an isolated chain.  In the presence of the interchain
interaction the string with the opposite sign of the dimerization
is created between the soliton and antisoliton.  It costs energy
proportional to the length of the string $l$, which results in a
linear potential between kinks,
\begin{equation}
\label{string} V(l) = \lambda {l}\;.
\end{equation}
The coefficient $\lambda$ measures the strength of the interchain
interaction and may be called the string tension.  The potential
tends to bind solitons and antisolitons into pairs.  The phase
transition now occurs at a finite temperature, when the thermally
excited soliton-antisoliton pairs dissociate.  This happens when
the pair size,
\begin{equation}
\label{pairsize}
R(T) = \frac{T}{\lambda}\;\;,
\end{equation}
becomes comparable to the distance between the thermally induced
kinks in an isolated chain,
\begin{equation}
\label{d(T)}
d(T) = \exp(\frac{\mu}{T})
\end{equation}
where $\mu$ is the kink excitation energy, which is of the order
of gap in electron (spin) excitation spectrum.  In this way for
weak interchain coupling, $\lambda \ll \mu$, we obtain
approximately
\begin{equation}
\label{Tc}
T_{c}(0) \sim \frac{\mu}{\ln\frac{{\mbox \large \mu}}
{{\mbox \large \lambda}}} \;\;.
\end{equation}

In the doped Peierls system, apart from the thermally induced
kinks, there exist also kinks induced by disorder, whose density
is proportional to the concentration of dopands.  The total
density of kinks is
\begin{equation}
n_{tot} = n_{therm}  + n_{dis} = e^{- \frac{\mu}{T}} + C x\;\;,
\end{equation}
where $C$ is some numerical factor of the order of $1$.  We can
now estimate the phase transition temperature $T_c(x)$ in the
same way as we did for a clean system, using $1 / n_{tot}$ as the
average distance between the kinks.  At small $x$, when the
density of the disorder-induced kinks $n_{dis}$ is much smaller
than the density of the thermally excited kinks $n_{therm}$, the
phase transition temperature decreases linearly with $x$,
\begin{equation}
\label{linear}
T_{c}(x) = T_{c}(0) \left( 1 - A x \right) \;\;,
\end{equation}
and the coefficient $A$ is large,
\begin{equation}
\label{slope}
A \sim \frac{\mu}
{\lambda \left( \ln \frac{\mu}{\lambda} \right)^2} \;\;.
\end{equation}
A more careful estimate of $A$ will be given in the next section.
If, on the other hand, $n_{dis} \gg n_{therm}$, the phase
transition temperature becomes inversely proportional to the
concentration,
\begin{equation}
\label{hyper}
T_{c}(x) = \frac{C}{x} \;\;.
\end{equation}
The interpretation of the last equation is, that at the phase
transition point the typical distance from kink to the nearest
end of odd segment, $\sim T / \lambda$, becomes comparable with
the average length of the segment $\sim 1 / x$.  This result is
not applicable at very large $x$: when an average length of the
segment becomes of the order of soliton size, $1 / x \sim \xi_0$,
there will be no ordering even at $T = 0$.

Because the density of the thermally induced kinks is
exponentially small at $T \ll \mu$, the disorder-induced kinks
begin to dominate at a very low doping concentration,
\begin{equation}
x_c \sim \frac{\lambda}{\mu} \ln(\frac{\mu}{\lambda}) \;\;,
\end{equation}
and the crossover between linear and $\frac{1}{x}$ decrease of
the transition temperature is very sharp for $\lambda \ll \mu$.

\section{Effective Ising model}

The model qualitatively discussed above, which describes the
statistical properties of kinks, induced thermally or by disorder
in the ensemble of weakly interacting finite chains, can be
formulated as an Ising-type model.  To this end we divide chains
into cells containing two bonds.  The sequence of bonds in each
cell can be either short-long or long-short (positive or negative
dimerization), which is decribed by the Ising variable $\sigma =
\pm 1$.  The uniformly dimerized lattice corresponds to the
ferromagnetic ordering of the Ising spins.  Kinks are domain
walls in Ising variables, {\em i.e.}, $\sigma_n = -1$ for $n$
less than some $m$ and $\sigma_n = +1$ for $n \geq m$, or {\em
vice versa}.  In this model one neglects the actual size of the
kink, which is reasonable as long as it is much smaller than the
average distance between the kinks.  The weak interaction between
different chains we treat in the mean-field
approximation.\cite{IPS} The energy of the chain containing $N$
Ising spins has the form,
\begin{equation}
\label{chain}
E_N^{\nu \nu^{\prime}} = \frac{\mu}{2}
\sum_{n = 1}^{N} \left( 1 - \sigma_{n} \sigma_{n+1} \right) -
h \sum_{n=1}^{N} \sigma_{n} -
\frac{\mu^{\prime}}{2} \left(
\nu \sigma_1 + \nu^{\prime} \sigma_N \right) + \mu^{\prime}
\;\;.
\end{equation}
Here the first term describes the energy cost of a kink (equal to
$\mu$), while the second term describes the interaction with the
mean field $h$ of neighbouring chains.  The third term is
introduced to model different kinds of finite chains in
(spin-)Peierls material.  It has a form of the interaction with
the ``magnetic field'' $h_L = \nu \mu^{\prime}$, applied at the
left end, and $h_R = \nu^{\prime} \mu^{\prime}$ at the right end.
Here $\nu$ and $\nu^{\prime}$ take values $\pm1$, which
corresponds to four kinds of finite (spin-)Peierls chains: even
segments in phase ($\nu = \nu^{\prime} = + 1$) and out of phase
($\nu = \nu^{\prime} = - 1$) with an average dimerization, and
odd segments with one end in phase and other out of phase ($\nu =
+ 1,\;\nu^{\prime} = - 1$ or {\em vice versa}).  In the latter
two cases, the opposite orientation of the magnetic field at
the chain ends creates a kink in the lowest energy configuration
of the Ising chain, provided that $\mu^{\prime} > \mu$.  In what
follows, we will restrict ourselves to the case of infinite
$\mu^{\prime}$.

The partition function of the Ising chain can be easily found by
means of the transfer matrix method.\cite{Kogut} For
$\nu^{\prime} = \nu$ the result is,
\begin{equation}
\label{pp}
Z_N^{\nu \nu} =
\frac{e^{- \nu {\bar h}}}{2}
\left[ (\lambda_+^{N+1} + \lambda_-^{N+1})
+ \nu \sin\psi (\lambda_+^{N+1} - \lambda_-^{N+1})
\right] \;,
\end{equation}
while for $\nu^{\prime} = - \nu$ one has,
\begin{equation}
\label{pm}
Z_N^{+ -} = Z_N^{- +} =
\frac{\cos\psi}{2}
(\lambda_+^{N+1} - \lambda_-^{N+1})\;.
\end{equation}
Here we denoted by $\lambda_{\pm}$ the eigenvalues of the
transfer matrix,
\begin{equation}
\lambda_{\pm} = \cosh{\bar h} \pm
\sqrt{\left(\sinh{\bar h}\right)^2 + e^{- 2 {\bar \mu}}} \;\;,
\end{equation}
and $\psi$ is defined by,
\begin{equation}
\sin\psi = \frac{\sinh{\bar h}}
{\sqrt{\left(\sinh{\bar h}\right)^2 + e^{- 2 {\bar \mu}}}} \;.
\end{equation}
Above we used the notations: ${\bar h} = \beta h$ and ${\bar \mu}
= \beta \mu$.

The averaging over disorder reduces in our model to the averaging
over $N$, $\nu$, and $\nu^{\prime}$.  For simplicity we assume
that for each $N$ all the four types of finite chains are
possible.  The distribution of the chain lengths $P_N$ for small
concentration $x$ of randomly positioned dopands ({\em e.g.} Zn
in CuGeO$_3$) is
\begin{equation}
P_N = \frac{1}{\langle N \rangle}
\exp \left( - \frac{N}{\langle N \rangle} \right) \;,
\end{equation}
where the average chain length is related to the concentration of
dopands by,
\begin{equation}
\langle N \rangle = \frac{2}{x}\;,
\end{equation}
(factor $2$ appears because the length of the unit cell in the
effective Ising model is equal to $2a$).  For small concentration
$x$ one can substitute the summation over $N$ by the integration.

The averaged free energy per unit cell is
\begin{equation}
\langle f \rangle = \frac{1}{4 \langle N \rangle}
\sum_{N,\nu,\nu^{\prime}} P_N
F_N^{\nu \nu^{\prime}}\;,
\end{equation}
where $F_N^{\nu \nu^{\prime}} = - T \ln Z_N^{\nu
\nu^{\prime}}$ is the free energy of a finite
chain. The mean-field value $h$ is proportional to the
averaged value of the Ising spin (the average dimerization),
\begin{equation}
\label{self}
h = \lambda \langle \sigma \rangle =
- \lambda \frac{\partial \langle f \rangle}{\partial h}\;,
\end{equation}
where $\lambda$ is the coupling constant.

The transition temperature, $T = T_c$, is obtained by soving
equation\cite{IPS}
\begin{equation}
- \lambda  \frac{\partial^2 \langle f \rangle}{\partial h^2}
(h = 0) = 1\;,
\end{equation}
Evaluation of the second derivative of the average free energy
gives
\begin{equation}
\label{tc}
\frac{\lambda e^{\bar{\mu}}}{T_c}
\left[
\frac{\gamma^2}{2} \zeta(2,\frac{\gamma}{2}) - 1
- 2 \epsilon \gamma
+ 4 \epsilon \gamma^3 \left( \beta(\gamma) - \frac{1}{2\gamma}
\right) \right] = 1\;,
\end{equation}
where $\zeta(z,q)$ is the Riemann zeta function,\cite{GR1}
\begin{equation}
\zeta(z,q) = \sum_{n=0}^{\infty}\frac{1}{(n + q)^z}\;,
\end{equation}
and $\beta(q)$ is,\cite{GR2}
\begin{equation}
\beta(q) =  \sum_{n=0}^{\infty}\frac{\;(-)^n}{n + q}\;.
\end{equation}
We also introduced the dimensionless parameter
\begin{equation}
\gamma = x e^{{\bar \mu}}\;,
\end{equation}
which is, roughly speaking, the ratio of the densities of the
disorder-induced and the thermally-induced kinks, and
\begin{equation}
\epsilon = \frac{1}{2}e^{{\bar \mu}}
\ln \left(\coth \frac{{\bar \mu}}{2} \right)\;.
\end{equation}
For $T \ll \mu$ $\epsilon$ is close to $1$.

If now $\gamma \ll 1$, {\em i.e.} the average distance between
the thermally induced kinks is much smaller than the average
chain size, the equation for $T_c(x)$ can be approximately
written as follows:
\begin{equation}
\frac{\lambda \exp \left(\frac{\mu}{T_c(x)}\right)}{T_c(x)}
\left( 1 - 2 \epsilon x
\exp\left( \frac{\mu}{T_c(x)}\right)\right) = 1\;.
\end{equation}
At $x = 0$ the critical temperature is determined by the
equation,
\begin{equation}
\exp\left(\frac{\mu}{T_c(0)}\right) = \frac{T_c(0)}{\lambda}\;,
\end{equation}
which coincides exactly with the condition $R(T_c(0)) =
d(T_c(0))$, which we used in the previous Section (see Eqs.
(\ref{pairsize}), (\ref{d(T)}), and (\ref{Tc})).
The magnitude of the linear slope of $T_c(x)$-curve at $x = 0$
(cf. Eqs.(\ref{linear}) and (\ref{slope})) is
\begin{equation}
\label{A}
A = - \frac{\frac{dT_c}{dx}(0)}{T_c(0)} =
\frac{2 \epsilon T_c(0)}{\lambda
\ln \left( \frac{e T_c(0)}{\lambda}\right)}\;.
\end{equation}

In the opposite limit $\gamma \gg 1$, when the disorder-induced
solitons dominate, we obtain from Eq.(\ref{tc}),
\begin{equation}
T_c(x) = \frac{\lambda}{6x}\;,
\end{equation}
which has to be compared with Eq.(\ref{hyper}) of the
previous Section.

Eq.(\ref{tc}) was solved numerically for the concentration
of dopands $x \leq 0.08$ and several values $\lambda$.  The
results are presented in Fig.~4. The value of $\mu$ for each
$\lambda$ was chosen to maintain $T_c(0) = 14K$. We see
that the crossover between the linear and $\frac{1}{x}$ behavior
of $T_c(x)$ is indeed very sharp.

\section{Conclusions}

The treatment carried above shows the important role played by
solitons in describing properties of both clean and doped
(spin-)Peierls systems.  For the electronic Peierls material
like polyacetilene, there are already a lot of experimental
evidences of the existence of kinks (although due to
a large kink energy they do not contribute significantly
to the thermodynamic properties of this material at room
temperature).

The mean-field soliton lattice solution was used to describe the
thermodynamics of the incommensurate phase of spin-Peierls
systems in strong magnetic field.\cite{BBK,Fujita,slat1,slat2} We
have shown, however, that solitons could strongly affect the
properties of spin-Peierls systems, both clean and doped, also at
$H=0$.  The small parameter $\lambda/\mu$ (where $\lambda$ is the
strength of the interchain interaction and $\mu$ is the kink
energy of the order of spin gap) is responsible for the sharp
suppression of $T_{c}$ by doping (cf.  Eqs.(\ref{slope}) and
(\ref{A})).  Qualitative estimates show, that the SP phase is
completely suppressed, when the average distance between
impurities becomes of the order of the coherence length of the SP
ordering $\xi$,
\begin{equation}
\xi\sim\frac{J_0}{E_g}a\;,
\end{equation}
or, in other words, when the average segment length
becomes of the order of the kink size.  For CuGeO$_3$,  $\xi \sim
8$--$10$ lattice spacings, so that the critical concentration for
the complete suppression of SP transition $x_c$ cannot, in any
case, exceed $\sim10$--$12\,$\%.  Experiments give a somewhat
smaller value: $x_c\sim7$--$8\,$\% for Zn and $x_c\sim2$--$3\,$\%
for Si.  The extra ``efficiency'' of Si, as compared to Zn, is,
most probably, related to the fact that Si, substituting Ge, is
located between two CuO$_2$-chains and thus influences two chains
simultaneously.  Therefore, two kinks are created per one Si and
only one kink per Zn.  Correspondingly, Si doping may be expected
to be two times more efficient in suppressing $T_{c}$ than the
doping by Zn.

An additional argument in favour of solitons induced in doped
(spin-)Peierls systems, comes from certain analogy between the
properties of doped CuGeO$_3$ and the behaviour of this material
at strong magnetic field\cite{buechner} (in which case the
soliton picture was checked experimentally\cite{slat1,slat2}).
The notion of solitons can, therefore, provide a unified approach
to this problems.

Yet another problem, appearing in the discussion of the doped
spin-Peierls systems, is the interplay between the SP and
antiferromagnetic ordering.  As was first observed in
Refs.\CITE{Zn2} and \CITE{Si2}, the two phases coexist in some
interval of the concentration of dopands.  Soliton picture
provides a natural explanation of this phenomenon\cite{KGM} As
was discussed above, the kinks created by doping carry
spin~$\frac12$.  They induce in their vicinitiy the
antiferromagnetic correlations, which decay at the length
scale~$\xi$ along the chain (see Fig.5).  The antiferromagnetic
microregions, created in this way, interact with each other,
giving finally the long-range magnetic order, but with a reduced
sublattice magnetization, spatially varying due to disorder.  It
is important that the spins of the separated kinks are free, so
that they can ajust to the mean field created by other spins.  In
other words, the disorder, inducing unpaired spins, does not
introduce frustration in the magnetic interaction.  Once the
coherence between the spins is established, the Neel temperature,
$T_N$, should grow, roughly speaking, linearly with the density
of the disorder-induced spins, {\em i.e.}, $T_N \propto x$ at
small $x$.  The detailed treatment of the interplay of the SP and
antiferromagnetic ordering will be given in a separate
publication.

A similar picture, but without an apparent use of the soliton
concept, in which the local antiferromagnetic correlations are
created in the regions where the lattice ajusts to impurities,
was suggested in Ref.\CITE{FTS}.  Creation of free spins by
nonmagnetic impurities in the rigidly dimerized chains was
considered in Ref.\CITE{MDR}.

Our model is, of course, oversimplified, as we assumed that the
phase transition into disordered state occurs entirely due to the
loss of coherence between the phases of the order parameter on
different chains.  In practice, the conventional spin and phonon
excitations have to be included to get a reasonable description
of the excitation spectrum, magnetic susceptibility and other
properties of spin-Peierls materials.  The aim of this
contribution, is, however, to show that the large fluctuations of
the order parameter (kinks), which are neglected in the
mean-filed approach, can be important in the
quasi-one-dimensional systems.

Summarizing, we can say that the concept of solitons, which is
well established for the electronic Peierls materials, is also
very useful for the discussion of spin-Peierls systems.  It gives
a natural framework for the description of the properties of
disordered Peierls systems and allows one to explain the
dependence of the spin-Peierls transition temperature on doping,
the interplay of the spin-Peierls and antiferromagnetic ordering,
the simultaneous influence of doping and magnetic field, etc.

\section*{Acknowledgments}

The authors are grateful to B.~B\"uchner, M.~T.~Figge,
W.~Geertsma, J.~Knoester and G.~Sawatzky for useful discussions.
This work was supported by the Dutch Foundation for Fundamental
Research of Matter (FOM).

\newpage

\section*{Figure captions}

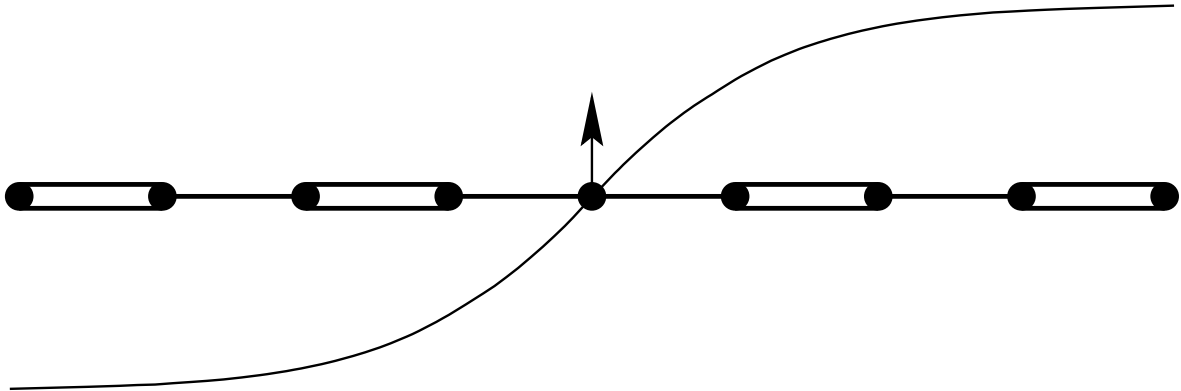
FIG.1.  Schematic view of the minimal energy lattice
configuration of an isolated finite (spin-)Peierls chain with
even (a) and odd (b) number of atoms (double and single lines
correspond, respectively, to short and long bonds).  The order
parameter (thin line) changes sign in odd chain, and one unpaired
spin (indicated by an arrow) is located near the kink.

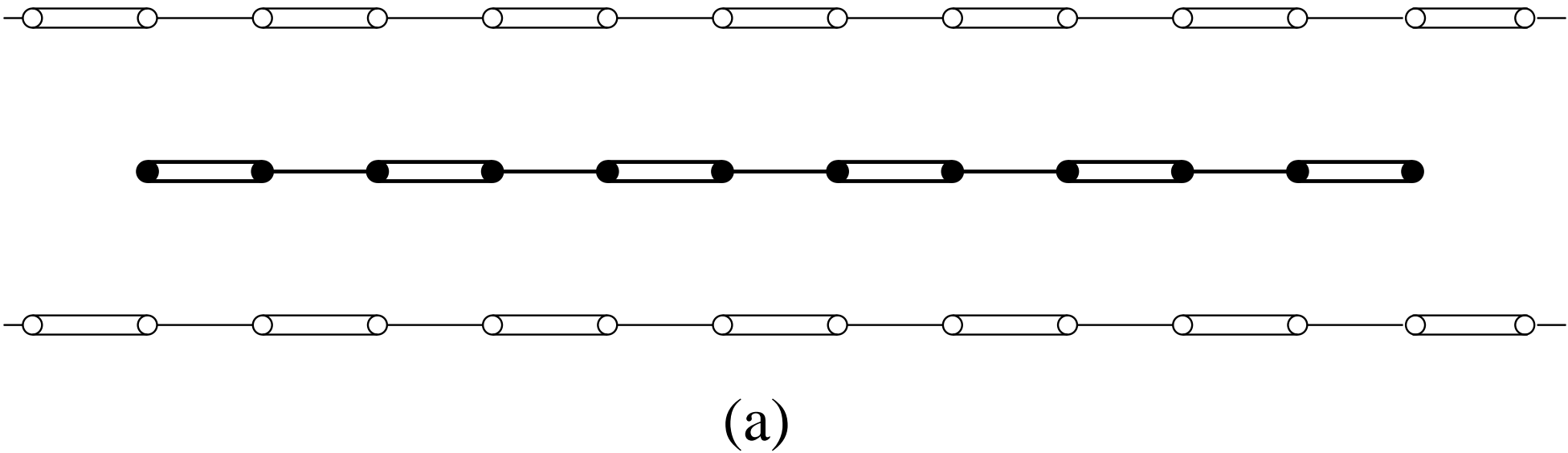
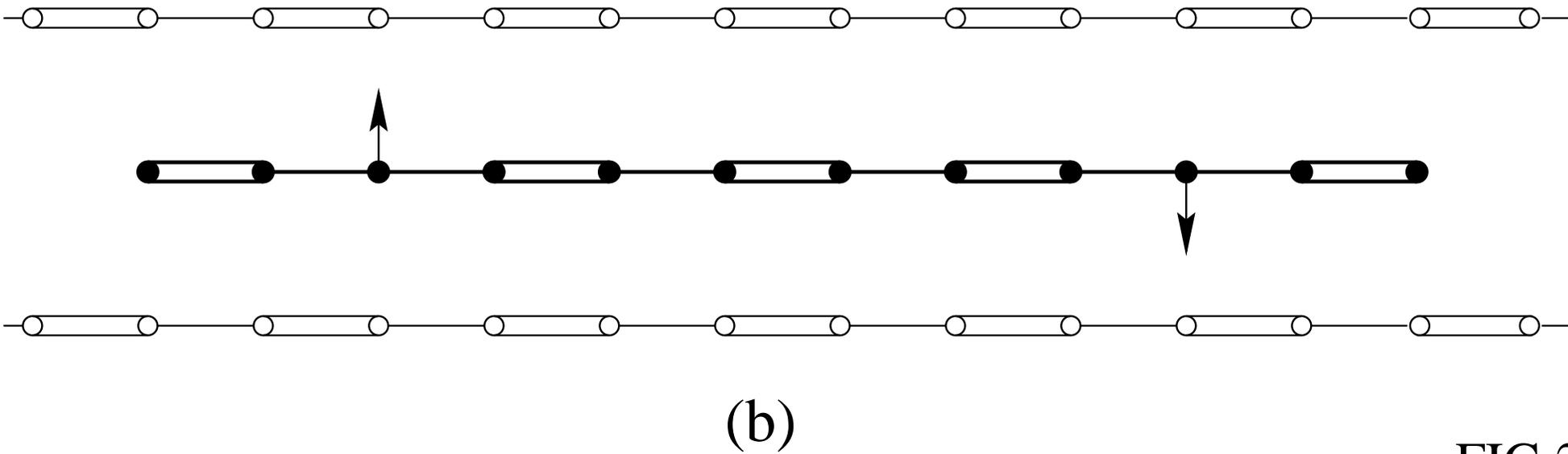
FIG.2.  Finite segment (thick lines) with even number of sites
out of phase with the surrounding chains (thin lines) creates a
string of unfavourable phase (a).  The string can be removed by
creating a kink and antikink near the chain ends (b).

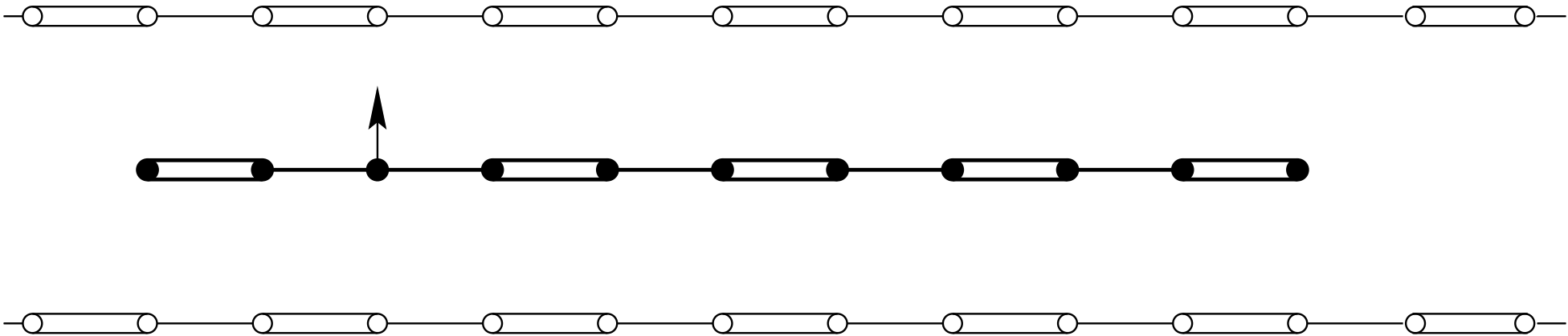
FIG.3. Kink in an odd segment moved to one of the chain ends to
minimize the interchain interaction energy.

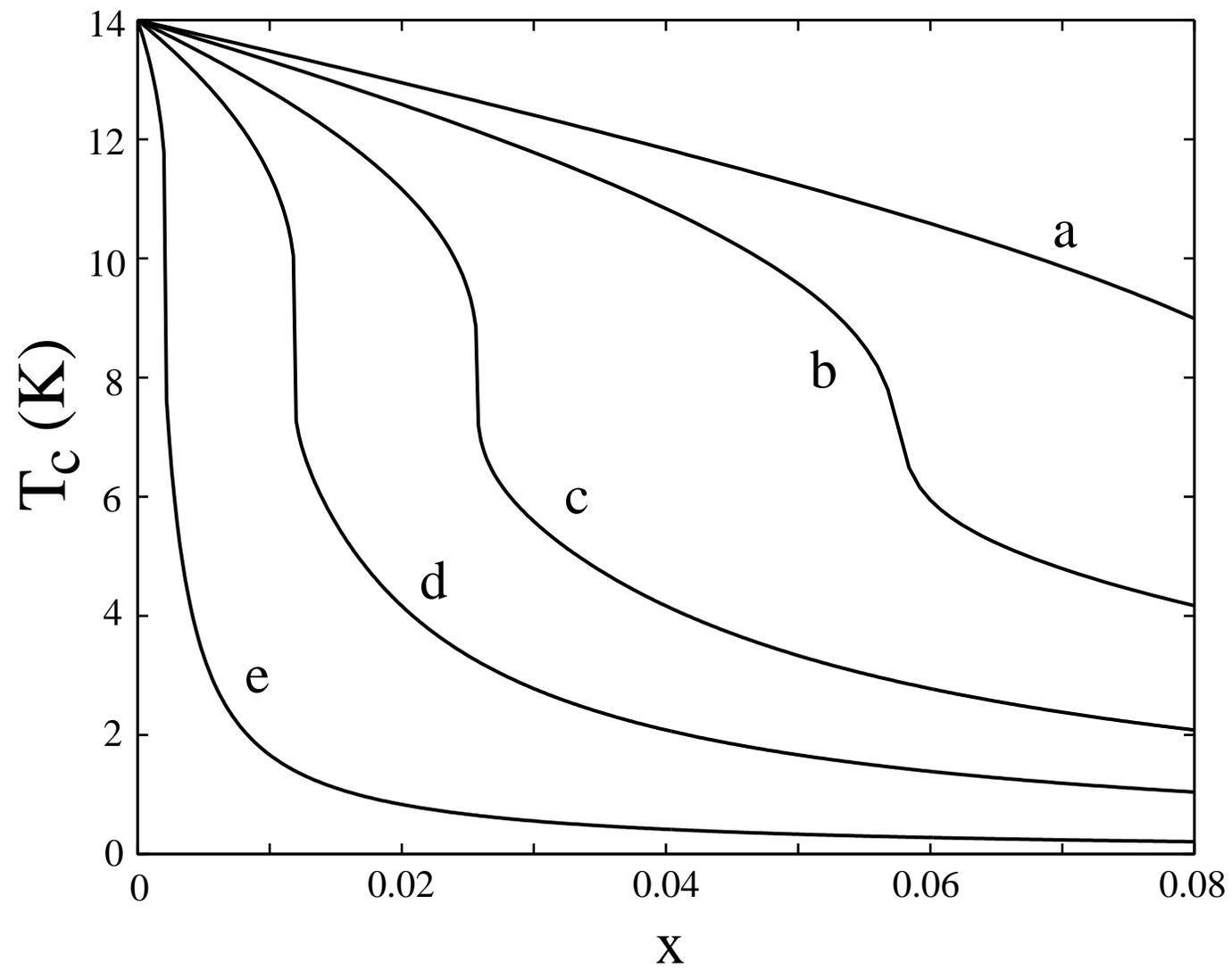
FIG.4.  Dependence of the (spin-)Peierls transition temperature
on concentration of dopands for $\lambda = 3,2,1,0.5,0.1$K (the
curves, respectively, (a) - (e)).

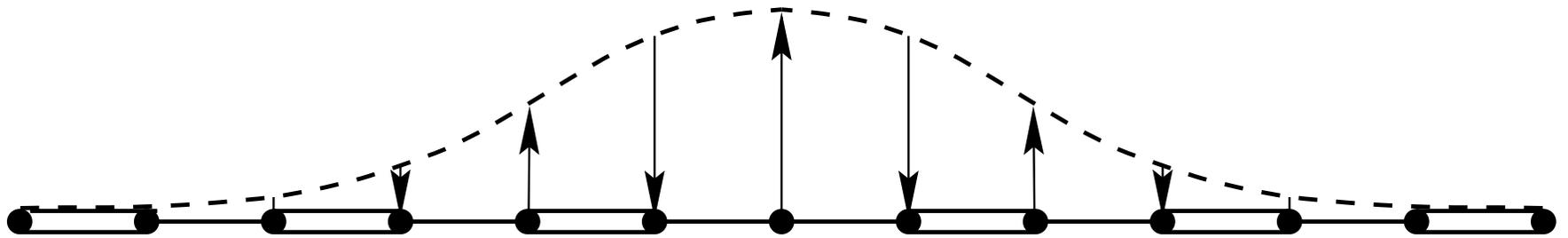
FIG.5. Antiferromagnetic correlations induced in the vicinity of
kink.

\end{document}